\documentclass[journal]{IEEEtran}
\usepackage{cite}	
\ifx\pdfoutput\undefined
\usepackage{graphicx}
\else
\usepackage[pdftex]{graphicx}
\fi
\usepackage{epsfig} 
\usepackage{psfrag}    
\usepackage{subfigure} 
\usepackage{url}       
\usepackage{amsmath}
\usepackage{array}
\usepackage{multirow}
\usepackage{threeparttable}
\usepackage{units}
\usepackage{booktabs}
\usepackage{xcolor}
\usepackage{amsmath,amsfonts,amssymb}
\usepackage{hyperref}
\usepackage{float}

\ifx\pdfoutput\undefined
\fi
\usepackage{threeparttable}
\usepackage{epstopdf}

\usepackage{graphicx}
\usepackage{dcolumn} 
\usepackage{graphics}
\begin{document}
\title{{\it{\small{Conference paper: 2019 IEEE International Electron Devices Meeting}}}\\
Challenges in Scaling-up the Control Interface of a Quantum Computer}
\author{{\bf D. J. Reilly}\\
 {\it Microsoft Quantum Sydney, The University of Sydney, NSW, 2006, Australia.}}
\maketitle
\begin{abstract}
Challenges at the quantum-classical interface are examined with the goal of architecting a scaled-up quantum computer comprising many thousands of qubits in the solid-state. Separating the distinct sub-systems of the interface that perform readout and control, general arguments are given for why distributing the  components of these sub-systems over significant distances and across large temperature gradients presents a major challenge to scaling-up the technology. Largely addressing these issues, an architecture for the interface that leverages cryo-CMOS circuits proximal to the quantum plane is motivated in addition to protocols that enable massively parallel readout of qubits via frequency multiplexing. 
\end{abstract}


\IEEEpeerreviewmaketitle

\section{Introduction}
\IEEEPARstart{R}{ealising} a quantum computer at the scale needed to address `real-world' problems is a formidable scientific, technical, and industrial challenge requiring new multi-disciplinary approaches to research, engineering, and workforce training over a sustained period. To date, research efforts spanning a variety of physical platforms have mostly focused on realising qubits (the fundamental building-blocks of quantum computers) that will likely soon yield control and readout fidelities sufficient to enable the use of error correcting protocols. This intense effort, underway now for more than two decades, has resulted in dramatic improvements in qubit coherence times and logic-gate fidelities in many systems \cite{Stajic1163}. As focus begins to shift to scaling-up the number of qubits in a quantum machine, new additional challenges, not apparent in today's present toy systems, are likely to emerge. 

The desire to predict these emergent `cliff-face' challenges in time to devise solutions motivates analysis of a quantum machine at a systems-level, including the scalability of the quantum-classical interface (QCI) that uses electrical signals to exchange information between a large number of qubits and the classical computer orchestrating a quantum algorithm \cite{DJR_NPJ,Lieven}. Indeed, the scalability of the interface and its related readout and control components is today a barrier to constructing quantum machines with large numbers of qubits, even in the so called  `NISQ-era'  \cite{Preskill2018quantumcomputingin}. This challenge of scaling the interface will likely also remain even when the fidelities of physical qubits are sufficient to enable the construction of logical qubits with extremely low error rates. 

Here, some of the well-known challenges that impact the scalability of the QCI are called-out and general arguments are made to suggest that separating qubits from their control systems can become problematic as the system is scaled-up. Careful consideration of these challenges suggests a solution based on the tight integration of control electronics and the qubit platform, with both operating at milli-kelvin temperatures. The present article provides context and motivation for the approach detailed in Ref. \cite{pauka2019cryogenic}.

\section{Scalability}
In quantum information processing the term `scalable' has long been used to describe the sense in which a large, complex machine can be constructed from fundamental building-blocks. At the qubit level, the word `scalable' often implies that the properties of single qubits remain the same, or even improve when the qubit is located in a multi-qubit system. That is, the ability to address a particular qubit does not change as the number of qubits is increased, or, that the readout and control fidelities or clock speed, in the worst case, remain the same as the number of qubits grows. 

Here, the term `scalability' is used to capture additional technical aspects that also need to be addressed in order to build-out a large, complex quantum machine. Examples include the architecture for managing input-output (IO) signals between qubits and the classical readout and control electronic sub-systems, with specific attention to system footprint, power dissipation, crosstalk, noise, and classical compute overhead. 

\section{The Quantum-Classical Interface (QCI)}
A scaled-up quantum computer will likely comprise a stack of sub-systems, with each layer of abstraction largely cloaking the layer beneath. Within this stack, the layer defining the QCI connects the platform-specific manipulation of physical qubits to the higher-level software tools needed to write and compile quantum algorithms. The QCI includes the electronic sub-systems of readout and control, such as data converters, amplifiers, signal sources, and digital logic responsible for generating and detecting analog (microwave) waveforms, sequencing and synchronising them, as well as the infrastructure that connects those signal paths to the physical qubit devices that encode quantum information. This second aspect comprises cabling, packaging, chip-interconnects, resonators, and on-chip routing and multiplexing approaches that together, constitute IO management. The QCI must also handle information exchange upwards in the stack, that is, between the classical compute platforms that essentially orchestrate the running of a quantum algorithm.\\

\section{Quantum Computers are Different}
A common refrain used to defend the scalability of quantum computing is to look to modern very large scale integration (VLSI) circuits, noting, for instance, that the processor in an iphone-X has 4.3 billion, 10 nm finFET transistors on-chip, but only requires a few hundred wires to exchange information with the outside world. Can qubits be tightly integrated in a similar way? Beyond the obvious technical differences between transistors and qubits, an important distinction is the way in which classical circuits `fan-out', enabling output signals from one logic gate to feed input signals for multiple gates (and vis-versa for fan-in). Rent's rule \cite{Rent} captures the degree of fanning, specifying the relationship between the number of external IO signals and the number of internal logic gates. 

Quantum circuits, however, do not fan-out in an analogous way: quantum information does not directly control the state of another qubit. Even for two-qubit logic gates, where qubits become entangled, the classical readout and control sub-systems are required to mediate the interactions in a switchable way as well as prepare and readout the state of the qubits. In addition, the `no-cloning' theorem in quantum mechanics \cite{Nielsen} holds that it is not possible to create an exact copy of a quantum state, ensuring that fan-out and fan-in of information must occur via the quantum-classical interface. This implies that the number of distinct IO signals must be at least as large as the number of qubits, although the physical resources used to propagate those signals can be used  more efficiently. Scaling-up qubits into large arrays will present an IO management challenge that likely requires heavy use of multiplexing approaches encompassing frequency, space, and time. The extent to which such techniques offer a ``free-lunch" over brute-force approaches will likely determine the scalability of the QCI. 

\section{Challengers at the Quantum-Classical Interface}
The nature of quantum circuits will require new approaches to address unique challenges posed by the fragility of qubits and the constraints of their extreme operating environments in comparison to complex classical electronics.  Central to the challenge of controlling a quantum computer is the fidelity of the quantum logic gates, which unlike classical digital computing, is exquisitely sensitive to noise or crosstalk across the system. In this sense quantum computers are similar to analog machines, at least until the system is of sufficient scale to enable quantum error correction protocols \cite{PhysRevLett.98.190504}. Moving forward from today's control setups, that are essentially complex physics experiments, to scalable architectures that enable the control of 1000s of qubits will require solutions to meet the following general issues:\\

\noindent $\bullet$ System Footprint \\
 The ENIAC was the world's first general purpose computer, it occupied an area of 167 m$^2$ and contained 5 million hand-soldered joints. Although ENIAC successfully accomplished it's mission, it was a large, distributed system built from macroscale components that had failure rates sufficiently high that the machine was non-functional approximately half the time. Today's quantum computers resemble ENIAC in a multitude of ways, comprising racks of electronic systems situated at room temperature and interfacing with qubit devices at milli-kelvin temperatures, at the bottom of a dilution refrigerator. As quantum machines are scaled-up, a dramatic shrinking in their footprint appears to be essential, following a similar path that saw classical machines evolve from room-sized, power-hungry beasts to tightly-integrated devices on a chip. 
 
{\em Is the size of the machine a fundamental barrier to scale-up?} 
In answering this question it is worth considering that there is no ``basic physics argument" that rules-out the construction of a large computer using vacuum tubes or even discrete transistors, and yet wiring up circuits in this way was abandoned for the most part as soon as integrated circuits took hold.

Putting aside the obvious convenience and advantages of a tightly-integrated system over a machine with components that span meters, from the perspective of signal propagation, a distributed system presents at least two additional challenges. Firstly, at the radio or microwave frequencies used for quantum control signals (MHz - GHz), the system size exceeds or is comparable to the wavelength. In this distributed regime, since currents and voltages are functions of space as well as time, it is necessary to work with controlled impedances, bringing additional constraints in power dissipation or increased footprint of the interconnects and matching circuits relative to systems in which the impedance is a free-parameter. The use of 50 $\Omega$ transmission lines, for instance, impacts the power dissipated by circuits that must drive a 50 $\Omega$ impedance. 

At the level of single qubits or even a handful, these aspects are hardly an impediment. For the case of scaled-up systems however, a distributed architecture that uses many parallel or branching transmission lines to span length-scales in excess of the signal wavelength introduces further hurdles since, differences in path-length lead to appreciable phase shifts in signals between different lines and interference between branches. Although these phase shifts are deterministic and can, in principle, be calibrated-out, this is a formidable challenge when synchronizing broadband signals that contain many different frequency components. The notion of synchronicity between signals on 1000s of different transmission lines must now account for their path length difference and spectral content (neglecting dispersion). This aspect is especially challenging for approaches that manipulate the state of qubits via the precise phase or amplitude of a microwave tone. 

A second consideration involves the time-of-flight latency for signals to travel the length of a cable between room temperature electronics (signal generators and acquisition systems) and an array of qubits operating a meter or so away, at deep cryogenic temperatures. Typical latencies, which are on the order of 10s of nanoseconds, are comparable to qubit gate times, but much shorter than typical readout times (100s of nanoseconds - microseconds for most solid-state qubits). In a scaled-up multi-qubit system however, where precise triggering across many readout and control sub-systems is required, it is not obvious that the signal latency can be ignored.  For instance, qubit stabilization protocols via feedback utilising a phase locked loop (PLL) can be problematic when the bandwidth is narrowed by time-of-flight latency. A path to reducing this latency may well be in the design of ultra-compact refrigeration systems, although additional constraints, such as an increase in thermal load from radiation and short cables are likely to appear. 

Again, these challenges for scaling qubits are different to those encountered in the synchronization of distributed classical computing components. In those classical systems, fan-in and fan-out of data between nodes essentially obeys Rent's rule \cite{Rent}, unlike large arrays of qubits that cannot be easily abstracted into ``black boxes" decoupled from their classical control interface.\\

\noindent$\bullet$ IO Management\\
Returning to the comparison with the ENIAC computer, the impact that component failure rate had on that  system's complexity (and scalability) is well documented. Interconnect and packaging solutions that make use of macroscopic connectors, cables, solder joints, etc, are challenging from this perspective in that their mechanical nature and the variability in their manufacture leads to failure rates or poor yield in excess of circuits that are lithographically defined and integrated on-chip or even printed circuit board.  A related challenge involves the footprint of wiring harnesses and connectors, which for high-frequency transmission lines also poses a challenge for scaling the number of IOs in a quantum computer \cite{CollessRSI}. The widely used SMA connector, for instance, measures 8 mm in diameter and even for  smaller `ganged' style connectors it is clear that this brute-force approach becomes increasingly challenging when the number of transmission lines exceeds a few hundred or so as the likelihood that the performance of at least one cable or connector has been degraded approaches certainty. Interconnect footprint is a particular burden for qubits that require the use of large magnetic fields (such as spin qubits or topological qubits) where the diameter of the magnet bore poses a further limitation to the footprint and density of connectors.  \\

\noindent$\bullet$ Heat and Power Dissipation \\
It is well known that the heat density of modern CPUs is the limiting factor in their clock speed. The origin of this power dissipation is mostly the charging and discharging of the transistors and interconnect capacitance with each clock cycle. In terms of power density and heat generation, quantum devices operating as qubits are hardly any different to the nanoscale transistors of today's CPUs. Although quantum computing at a deeply abstract level can be considered dissipation-less, as soon as real voltages, gate-capacitances, interconnects and classical control systems are considered, dissipation becomes a reality. As it is for classical nanoscale transistors, removing the heat generated by the dynamic operation of qubits, irrespective of the location or details of the their classical control interface, is a potential barrier to scale-up. This is a significant issue for qubit platforms that require large voltage (or current) pulses to control qubits, since power dissipation goes as the square of the voltage.

This heat problem is compounded by the need for a cryogenic environment to operate the qubits, since extremely limited cooling power is available  at these temperatures. Qubit platforms operating below 1 kelvin are the most constrained in this regard since the typical cooling power of a commercially available dilution fridge is of order 1 mW at 100 mK. For systems that can function at elevated temperatures, the 4 kelvin environment can deliver cooling powers measured in Watts or even kiloWatts when large quantities of liquid helium are used, although in this later regime the limited thermal conductivity of materials still presents a challenge for heat removal. 

The challenge of removing heat, whilst a major constraint on the performance of room temperature electronics, is so dramatic at cryogenic temperatures that it likely limits the extent to which active circuits can be used to address the challenges of system integration and scale-up. This is true even when entirely ignoring the efficiency of the cooling system in terms of ``wall-plug power". The tiny cooling power available at the qubit plane suggests then that the only viable approach is to locate the classical sub-systems of the control interface at elevated temperatures, where they are able to dissipate significant power. 

A competing constraint enters however, when considering the power dissipation and thermal conductivity of the cables and interconnects that must then be used to connect the warm control interface to the cold qubits. This issue is particularly significant when using transmission lines that incorporate attenuators (often used today to limit the amount of thermal noise propagating to the qubits from higher-temperature electronics). Attenuators are also used to thermalise cryogenic transmission lines by enabling a contact between the refrigerator and the inner-conductor of a cable. These attenuators however, also lead to resistive losses in control signals, generating substantial heat as the number of transmission lines is increased for controlling a large number of qubits (via brute force). In fact, even without adding attenuators the losses from the bare cables cannot be neglected at microwave frequencies. Scope to reduce these losses is limited since, for a normal metal the Weidemann-Franz law requires that increasing the electrical conductivity also increases the thermal conductivity, increasing the direct heatload. This is a particularly serious problem for qubit technologies that require large, time-varying voltages (10s of mVs) or currents (amps) at the bottom of the dilution refrigerator. 

The above arguments often lead to the conclusion that superconducting transmission lines should be deployed where possible to avoid cable attenuation. It is worth noting however, that the use of superconductors (which still have electrical losses at microwave frequencies and non-zero thermal conductivities) does not overcome the need to drive the cable impedance, that is, power is still required to charge the cable capacitance and inductance, and this power is eventually dissipated. Superconducting transmission lines can help by limiting the heat dissipated by the cable, but they do not address the power dissipated by the signal source, termination, or amplifier needed to drive the cable. In this sense, they do not suddenly solve the dilemma presented by positioning the control electronics at temperature stages above the qubit devices. \\

\noindent$\bullet$ Noise, Crosstalk, and Interference\\
Unlike today's classical computers that process digital information, qubits are essentially analog devices. Considering also the typical energy scales that characterise quantum devices (equivalent to temperatures of just a few kelvin), these aspects make qubits exquisitely sensitive to all forms of noise in the control signals themselves as well as noise originating from the qubit environment, ultimately leading to decoherence. In the case of the readout sub-system, similar constraints arise from the need for ultra-low noise amplifiers that can only be realised by cooling to cryogenic temperatures \cite{Macklin307}.
Examining the noise performance of the classical interface sub-systems, again the distributed nature of the control platform brings challenges to scale-up. For instance, where room temperature electronics is used to readout and control qubits at the bottom of dilution fridge, long cables increase the cross-section available for coupling interference and require additional shielding or approaches that minimize the effective loop available for pickup - a challenge in a constrained space like a cryostat. In a scaled-up multi-qubit control system, mutual inductance and capacitive coupling also lead to electromagnetic crosstalk between control and readout signals, degrading qubit fidelities and also leading to correlated errors across multiple qubits. Although this form of crosstalk is, in principle, deterministic, nulling its effect brings significant overhead in the operation of a scaled-up control platform.

Further, with the control electronics at room temperature, situated outside the cryogenic environment, thermal noise on control signals is a constant challenge to mitigate. As discussed above, thermal noise can be suppressed via the use of attenuators, but these also attenuate the signal and lead to additional power dissipation. For readout, thermal noise requires at least the first stage of amplification to be cooled to cryogenic temperatures in order to realise even modest readout fidelities.\\

\noindent$\bullet$ Slew Rate, Bandwidth, and Rise-time \\
With many qubit platforms controlled by pulse sequences that contain frequency components in the rf or microwave domain, wideband transmission lines, connectors, and packaging solutions are needed to connect waveform generators to quantum devices. Again, with the generator electronics situated at room temperature, a meter away from the milli-kelvin stage of a dilution refrigerator, the cable impedance and its attenuation can greatly limit the high-frequency performance of the systems. For example, consider qubit control protocols \cite{hybridqubit} that require square pulse with  rise-times of order 100 picosecond  and amplitude in the milli-volt range. Generating these at room temperature, before the signal is attenuated by at least 20 dB in propagating down the fridge, requires a generator with a slew-rate of $\sim$ 1V / 100 ps. Achieving such performance, which typically involves an additional wideband amplification stage in the waveform generator to drive the cable impedance, also adds significant noise. It is worth comparing this distributed setup to the performance of a standard integrated CMOS circuit, where the on-chip slew-rate is set by the $RC$ time of the transistors and interconnects. For modern technology nodes, on-chip rise-times  shorter than 10 picoseconds for milli-volt pulses can easily be met.

\section{A Way Forward: A Cryogenic Quantum - Classical Interface}
Many of the challenges to scaling, discussed above, appear to be largely addressed by dramatically shrinking the distributed control system and implementing it as a series of tightly packaged integrated circuits or components in close proximity to the qubits at cryogenic temperatures. The one challenge that stands out as an exception to this approach however, is the power dissipation problem: at the milli-kelvin temperature of the qubits the available cooling power is only 100 microwatts or so. If an ultra-low power technology can be deployed at cryogenic temperatures, many of the barriers to scaling the control interface become surmountable. For instance, on-chip control schemes immediately address the issue of footprint, bringing the length-scale of the circuit well below the wavelength of the signals used for control and readout. Time-of-flight signal latency is also obviously reduced by locating aspects of the control interface in close proximity to the qubit plane. This paradigm shift, from a large distributed system to on-chip tightly integrated circuit blocks, opens the possibility of deploying modern approaches to packaging for IO and interconnect management. Chip stacking and techniques based on controlled collapse chip connections (C4), for instance, can dramatically scale-up the density and reliability of IOs in comparison to macroscale wiring harnesses and connector solutions.\\
\begin{figure*}[!t]
\center
\includegraphics[scale=1]{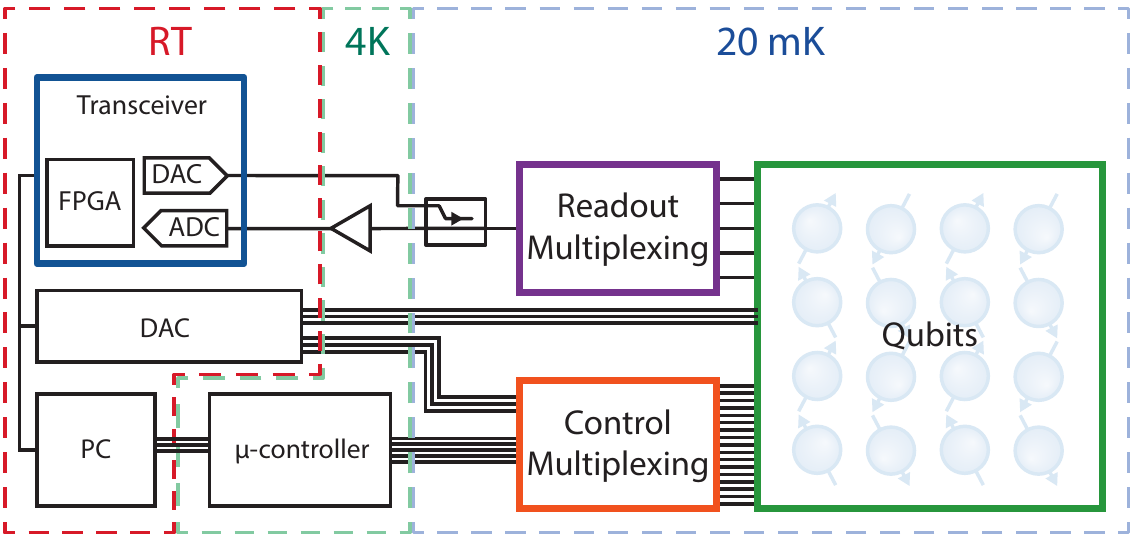}
\caption{\label{}High-level schematic of the quantum-classical interface that leverages ultra-low power cryo-CMOS circuits for control signal generation. The control MUX block handles IO management, reducing the need for many high-bandwidth cables to higher temperature stages, and enables dense IO via chip-stacking techniques \cite{pauka2019cryogenic}. For readout, signals are generated using direct digital synthesis (DDS), amplified at low temperatures and acquired by a high-speed ADC, operating at room temperature. A micro-controller, operating at 4K coordinates triggering, clocking, and command buffering.}
\end{figure*}

{\it A Scalable Control Approach based on Cryo-CMOS}\\

Circuits constructed from cryogenic-CMOS \cite{Extreme_electronics,8424046,8123682} appear well-placed to address many of the challenges that presently limit the scaling-up of the QCI. Again, the major hurdle to adopting cryo-CMOS  for control, however, is the power dissipation, which with the very limited cooling power available at the qubit plane, can be prohibitively high. Modern CPUs, for instance, dissipate several 10s of Watts, four orders-of-magnitude above the cooling power typically available at 100 mK. 

The key challenge of power dissipation in cryo-CMOS circuits is a relatively straightforward problem to at least define. For cryogenic operation it is possible to largely neglect static power dissipation that stems from transistor leakage, leaving active power alone, given by $P = CV^2f$. Here $C$ is the total capacitance comprising the transistors, parasitic, and interconnect contributions, the voltage $V$ can be decomposed into what is needed to switch a CMOS transistor as well as what is needed to control a qubit, and $f$ is the clock frequency. Assuming that the voltage needed at the qubit plane is essentially fixed by the physics of the qubit platform, reducing power dissipation can only proceed via a reduction in the number of transistors in a control circuit, their frequency of operation, or the capacitance of the transistors and interconnects. 

With the problem then defined as, how to architect an integrated control platform with absolute minimal power dissipation, the figure below captures a broad approach to a solution based on cryo-CMOS. The details of this approach are described in Ref. \cite{pauka2019cryogenic}.

Turning to the challenge of scalable readout, its worth first drawing attention to the difference between sensing the state of a qubit and controlling it. Similar to many engineered systems, the act of measuring the state of a system with a sensor is generally easier (requiring less energy) than controlling it with an actuator. This holds true for reading out qubits where detecting the phase or amplitude of an extremely low power microwave tone can determine the state of a qubit \cite{Reilly:2007ig}. The efficiency of the readout scheme can be extended then by the use of frequency multiplexing \cite{Hornibrook_APL}, encoding each qubit to a unique frequency transmitted on a single transmission line and cryogenic amplification chain, as shown in the figure above. 

\section{Conclusion}
In summary, a number of well-known challenges for scaling-up the quantum-classical interface of a quantum computer have been described. Many of these challenges stem from the large, distributed nature of present approaches that involve many sub-systems interconnected across large distances and vast temperature gradients.  The key barrier to constructing control systems that are tightly integrated with the qubit plane however, is the added power dissipation that an integrated platform would bring at cryogenic temperatures. An architecture that mostly addresses these challenges has been briefly described, tightly integrating the control and readout modules in close proximity to the qubits, operating at milli-Kelvin temperatures. Detailed analysis and cryo-CMOS prototyping \cite{yang2019cryocmos,pauka2019characterising} suggests that power dissipation can, in fact, be kept sufficiently low that the control and readout of 1000s of topological qubits or similar quantum device structures appears possible \cite{pauka2019cryogenic}.
\section{Acknowledgements}
This research was supported by Microsoft Corporation (Quantum Sydney) and the Australian Research Council Centre of Excellence Scheme (Grant No. EQuS CE110001013). The author wishes to thank S. Pauka for artwork, and K. Das, A. Moini, Y. Yang, J. Knoblaunch, M. Cassidy, and E. Hurbi for useful discussions.

 \bibliographystyle{IEEEtran}

\end{document}